\documentclass[10pt]{iopart}

\usepackage[cp1251]{inputenc}
\usepackage{iopams}
\usepackage{graphicx,epsfig}
\begin{document}

\title[Klein-Gordon]{The Klein-Gordon-Fock theory as a one-particle relativistic quantum mechanics}

\author{N. L. Chuprikov}

\address{Tomsk State Pedagogical University, 634041, Tomsk, Russia}
\ead{chnl@tspu.edu.ru} \vspace{10pt}

\begin{abstract}
The Klein-Gordon-Fock (KGF) theory is presented as a single-particle relativistic quantum theory. A Schrodinger-type
equation for a single-component wave function is derived. In addition to the KGF Hamiltonian, operators for momentum and
orbital angular momentum for a scalar boson are identified, which act within the space of single-component wave functions.
This theory is applicable to both massive and massless scalar bosons.

{\bf Key words}: Klein-Gordon-Fock equation; single-particle formulation.
\end{abstract}

\newcommand{\ppp}{\mbox{\hspace{5mm}}}
\newcommand{\ooo}{\mbox{\hspace{3mm}}}
\newcommand{\ooa}{\mbox{\hspace{1mm}}}
\newcommand{\ppd}{\mbox{\hspace{18mm}}}
\newcommand{\ppt}{\mbox{\hspace{34mm}}}
\newcommand{\ppo}{\mbox{\hspace{10mm}}}
\newcommand{\lcom}{\lambda\hspace{-1mm}\bar{}\hspace{1mm}}

\section{Introduction}

As is well known, the Klein-Gordon-Fock (KGF) equation and the Dirac equation were originally conceived by their authors
as relativistic generalizations of the corresponding non-relativistic single-particle equations of quantum mechanics (QM).
However, both equations proved to be fundamentally different from their non-relativistic counterparts -- the Schrodinger
equation and the Pauli equation -- and for many years, attempts to construct a consistent single-particle quantum
formulation for them were unsuccessful. Consequently, a widespread view emerged that such a formulation was impossible in
principle, and both equations came to be regarded as describing specific fields: a scalar boson field with spin $s=0$ and
a fermion field with spin $s=1/2$.

This situation changed when a quantum-mechanical single-particle formulation of the Dirac equation was presented in
\cite{Chu}. The main lesson to follow from this approach is that the equantion for the wave function of the Dirac particle
satisfies two key properties of the Pauli equation: (\i) it has the Scr\"{o}dinger-like (Hamiltonian) form; (\i\i) the
number of components of this wave function is equal to the number of internal degrees of freedom of the Dirac particle.

As regards the Klein-Gordon-Fock equation, it satisfies condition (ii), but does not meet the first condition. It lacks a
Hamiltonian form, and our goal is to develop a single-particle quantum-mechanical formulation of the Klein-Gordon-Fock
theory that satisfies both of these requirements.

An important step in solving this problem was made in the work of Feshbach and Villars \cite{Fesh}. To write the KGF
equation in a Hamiltonian form, they proposed considering the first time derivative of the unknown function as a new wave
function in addition to the original one. Although their proposed ``Hamiltonian form'' of the KGF equation is not actually
such (since it is written in terms of a non-Hermitian Hamiltonian), they set the right direction in the search for a  QM
form of the KGF equation.

\section{Background}\label{initial}

In quantum field theory (see, for example, \cite{Grei}), the KGF equation and the corresponding energy-momentum tensor are
obtained from the variational principle
\begin{eqnarray} \label{300}
\fl \delta \int\mathcal{L}\left(\Psi,\frac{\partial\Psi}{\partial x^\mu},\Psi^*,\frac{\partial\Psi^*}{\partial
x_\mu}\right) d^4x=0,
\end{eqnarray}
where the Lagrange density is
\begin{eqnarray} \label{301}
\fl \mathcal{L}=\frac{c^2}{2E_{tot}}\left[(p_\mu\Psi)^*(p^\mu\Psi)-m^2c^2\Psi^*\Psi\right]=
\frac{c^2}{2E_{tot}}\left[g^{\mu\nu}(p_\mu\Psi)^*(p_\nu\Psi)-m^2c^2\Psi^*\Psi\right];
\end{eqnarray}
$g_{\mu_\nu}$ is a metric tensor whose non-zero elements are $g_{00}=-g_{11}=-g_{22}=-g_{33}=1$;
\begin{eqnarray*}
\fl p^\mu=\left\{\frac{D_t}{c} ,\mathbf{p}\right\}\equiv i\hbar\left\{\frac{\partial}{c\partial
t},-\boldsymbol{\nabla}\right\};
\end{eqnarray*}
the Greek indices take the values $0,1,2,3$; for identical indices the Einstein summation rule is assumed. Note that,
unlike the usually used constant $1/2m$ (see for example \cite{Grei}), in the Lagrangian (\ref{301}) we employ the
constant $c^2/(2E_{tot})$, where $E_{tot}$ is the conserved total energy of the scalar boson's field (see (\ref{305}));
within our approach, the particle mass $m$ can be zero.

So, the KGF equation is
\begin{eqnarray} \label{1}
\fl (p_\mu p^\mu-m^2c^2)\Psi\equiv\left(D_t^2-c^2 \mathbf{p}^2-m^2 c^4\right)\Psi=0.
\end{eqnarray}
From this equation it follows the continuity equation $\partial_\mu j^\mu=0$, where
\begin{eqnarray} \label{302}
\fl j^\mu=\frac{ie\hbar}{2m}\left(\Psi\partial^\mu\Psi^*-\Psi^*\partial^\mu\Psi\right) \equiv
\{c\rho,\mathbf{j}\};\nonumber
\\ \fl \rho=\frac{ie\hbar}{2mc^2}\left(\Psi\frac{\partial\Psi^*}{\partial t}-\Psi^*\frac{\partial\Psi}{\partial t}\right),\ooo
\mathbf{j}=\frac{ie\hbar}{2mc^2}(\Psi^*\boldsymbol{\nabla}\Psi-\Psi\boldsymbol{\nabla}\Psi^*).
\end{eqnarray}
Since the function $\rho$ in this Lorentz-covariant continuity equation can take negative values, it is interpreted as
charge density (rather than probability density), and the three-dimensional vector $\mathbf{j}$ as electric current
density. Consequently, it is generally accepted that the Klein-Gordon-Fock equation cannot be formulated as a
quantum-mechanical equation.

There are also four conditions $\partial_\mu T^{\mu\nu}=0$ for the energy-momentum tensor $T^{\mu\nu}$:
\begin{eqnarray*}
\fl T^{\mu\nu} =\frac{\partial\mathcal{L}}{\partial(\partial_\mu\Psi^*)}\partial^\nu\Psi^*
+\frac{\partial\mathcal{L}}{\partial(\partial_\mu\Psi)}\partial^\nu\Psi -g^{\mu\nu}\mathcal{L},
\end{eqnarray*}
Taking into account Exp. (\ref{301}), we obtain
\begin{eqnarray*}
\fl T^{\mu\nu} =\frac{c^2}{2E_{tot}}\Big[\hbar^2(\partial^\mu\Psi\partial^\nu\Psi^*+\partial^\mu\Psi^*\partial^\nu\Psi) -
\delta^{\mu\nu}\left[\hbar^2(\partial_\alpha\Psi)^*(\partial^\alpha\Psi)-m^2c^2\Psi^*\Psi\right]\Big]
\end{eqnarray*}
Thus, in addition to the charge density $\rho$, there are also the energy density $T^{0 0}$ of the scalar boson field and
its momentum densities $P_n=T^{0 n}/c$ $(n=1,2,3)$:
\begin{eqnarray} \label{304}
\fl T^{00} =\frac{1}{2E_{tot}}\left[h^2\frac{\partial\Psi^*}{\partial t}\frac{\partial\Psi}{\partial t}
+c^2\hbar^2(\boldsymbol{\nabla}\Psi)^*(\boldsymbol{\nabla}\Psi)+m^2c^4\Psi^*\Psi \right],\nonumber\\ \fl T^{0n}
=-\frac{c\hbar^2}{2E_{tot}}\left(\frac{\partial\Psi}{\partial t}\frac{\partial\Psi^*}{\partial x_n} +
\frac{\partial\Psi^*}{\partial t}\frac{\partial\Psi}{\partial x_n}\right).
\end{eqnarray}

They lead to the conserved quantities -- the total energy $E_{tot}$ and total momentum $\mathbf{P}_{tot}$ of the scalar
boson field:
\begin{eqnarray} \label{305}
\fl E_{tot}=\int_{\mathbb{R}^3} T^{00}(\mathbf{r},t)d^3\mathbf{r},\ppp (P_n)_{tot}=\frac{1}{c}\int_{\mathbb{R}^3}
T^{0n}(\mathbf{r},t)d^3\mathbf{r},\ooo (n=1,2,3).
\end{eqnarray}

\section{A Hamiltonian pseudo-quantum-mechanical form of the KGF equation}\label{hamilton}

Note that Eq. (\ref{1}) can be written in the form
\begin{eqnarray} \label{1a}
\fl \left[D_t^2-(mc^2-icp)(mc^2+icp)\right]\Psi=0,
\end{eqnarray}
where $p=\sqrt{(\mathbf{p}\cdot\mathbf{p})}=-i\hbar\sqrt{(\boldsymbol{\nabla}\cdot\boldsymbol{\nabla})}$. This equation
prompts us to introduce two new functions $\phi_1$ and $\phi_2$ such that
\begin{eqnarray}\label{1aa}
\fl D_t\Psi=\epsilon(\phi_1-i\phi_2),\ooo (mc^2+icp)\Psi=\epsilon(\phi_1+i\phi_2),
\end{eqnarray}
where $\epsilon$ is an arbitrary real constant. Substituting these expressions into Eq. (\ref{1a}) and omitting the
constant $\epsilon$, we reduce this equation to the form
\begin{eqnarray*}
\fl D_t(\phi_1-i\phi_2)-(mc^2-icp)(\phi_1+i\phi_2)\\
\fl =D_t\phi_1-mc^2\phi_1-cp\phi_2+i\left(-D_t\phi_2-mc^2\phi_2+icp\phi_1\right)=0.
\end{eqnarray*}
It is evident that Eq. (\ref{1a}) for $\Psi$ is equivalent to the system of equations for $\phi_1$ and $\phi_2$
\begin{eqnarray} \label{3}
\fl i\hbar\frac{\partial\Phi}{\partial t}=H\Phi;\ppp H=\left(
\begin{array}{cc}
mc^2 & cp \\
cp & -mc^2
\end{array} \right),\ooo \Phi=\left(
\begin{array}{cc}
\phi_1 \\
\phi_2
\end{array} \right)
\end{eqnarray}
As is seen, this system of differential equations satisfies requirement (i) but violates requirement (ii). We shall refer
to it as an equation written in pseudo-quantum-mechanical Hamiltonian form. We will assume that the solutions of Eq.
(\ref{3}) form the space $L^2(\mathbb{R}^3)$ in which the scalar product for any two spinors $\Phi_1(\vec{r},t)$ and
$\Phi_2(\vec{r},t)$ is defined as $\langle\Phi_1|\Phi_2\rangle= \int\Phi_1^\dag(\vec{r},t)\Phi_2(\vec{r},t) d^3\vec{r}$.

From (\ref{3}) it follows the continuity equation
\begin{eqnarray} \label{4}
\fl \frac{\partial}{\partial t}(\Phi^\dag\Phi)\equiv\frac{\partial}{\partial t} \left(|\phi_1|^2 + |\phi_2|^2\right)=
\frac{c}{i\hbar}\left(\phi_1^*p\phi_2-\phi_1 p\phi_2^*+\phi_2^*p\phi_1-\phi_2 p\phi_1^*\right).
\end{eqnarray}
And since the operator $p$ is Hermitian, this means that
\begin{eqnarray} \label{4a}
\fl \frac{d}{d t}\langle\Phi|\Phi\rangle = \frac{c}{i\hbar}\left[\langle\phi_1|p\phi_2\rangle-\langle
p\phi_2|\phi_1\rangle+\langle\phi_2|p\phi_1\rangle -\langle p\phi_1|\phi_2\rangle\right]=0.
\end{eqnarray}
In the following, we assume that $\langle\Phi|\Phi\rangle =1$.

\section{Transition to a one-component wave function}\label{one}

Since a spin-0 particle possesses only one internal degree of freedom, its state as a quantum particle must be described
by a single-component wave function. Therefore, by analogy with the Dirac particle (see \cite{Chu}), we assume that for
any spinor $\Phi$ there exist an angular operator $\eta$ and a normalized function $f \in L^2(\mathbb{R}^3)$ such that
\begin{eqnarray} \label{5}
\fl \phi_1 = \cos\hat{\eta} f,\ppp\phi_2 = \sin\hat{\eta} f.
\end{eqnarray}
Substituting these expressions into (\ref{3}), we obtain
\begin{eqnarray} \label{7}
\fl [D_t\cos\eta-mc^2\cos\eta-cp\sin\eta]f=0,\ppp [D_t\sin\eta-cp\cos\eta+mc^2\sin\eta]f=0
\end{eqnarray}

According to (\ref{3}), the operator $D_t\equiv i\hbar\partial/\partial t$ acts in the space of spinors $\Phi$ as the
matrix operator $H$. But its action in the space of one-component functions $f$ must obviously be different. The form of
this operator is determined by the system of Eqs. (\ref{7}), which should be viewed as a system of equations for the
operators $i\hbar\partial/\partial t$ and $\eta$, and not for the function $f$. In other words, these equations must hold
for an arbitrary function $f$, and therefore the following compatibility conditions must be satisfied:
\begin{eqnarray} \label{8}
\fl \tan\eta=\frac{D_t-mc^2}{cp}=\frac{cp}{D_t+mc^2},
\end{eqnarray}
from which it follows that the operator $D_t$ must satisfy the equation
\begin{eqnarray} \label{8a}
\fl (D_t)^2=c^2p^2+m^2c^4.
\end{eqnarray}
This equation has two roots:
\begin{eqnarray} \label{9}
\fl D_t^{(\pm)}=\pm c\sqrt{p^2+m^2c^2};
\end{eqnarray}
where the operator $D_t^{(+)}$ is bounded below, and the operator $D_t^{(-)}$ is bounded above.

As in the work \cite{Chu}, it is convenient to introduce operators
\begin{eqnarray}\label{9m}
\fl M=\frac{1}{2}\left[\sqrt{m+\mathcal{M}^2}+m\right],\ooo \mu=\frac{1}{2}\left[\sqrt{m+\mathcal{M}^2}-m\right];\ppp
\mathcal{M}=\frac{p}{c}.
\end{eqnarray}
It is ivident that
\begin{eqnarray*}
\fl M+\mu=m\sqrt{1+\xi^2},\ppp M-\mu=m,\ppp 2c\sqrt{M\mu}=p.
\end{eqnarray*}
Thus,
\begin{eqnarray} \label{9a}
\fl D_t^{(\pm)}=\pm c\sqrt{p^2+m^2c^2}=\pm (M+\mu)c^2
\end{eqnarray}
and, hence
\begin{eqnarray}\label{11}
\fl \sin\eta_{(+)}=-\cos\eta_{(-)}=\sqrt{\frac{\mu}{M+\mu}},\ppp \cos\eta_{(+)}=\sin\eta_{(-)}= \sqrt{\frac{M}{M+\mu}}.
\end{eqnarray}
From now on we will omit the index $(+)$ of the angular operator $\eta_{(+)}$.

\section{Quantum-mechanical Hamiltonian form of the KGF equation} \label{quant}

Note that Eq. (\ref{3}) implies an equality
\begin{eqnarray}\label{12}
\fl i\hbar \left\langle\Phi_1\bigg|\frac{\partial \Phi}{\partial t}\right\rangle=\langle\Phi_1|H\Phi\rangle= \langle
H\Phi_1|\Phi\rangle,
\end{eqnarray}
which must be satisfied for any spinors $\Phi_1$ and $\Phi$, including spinors
\begin{eqnarray}\label{13}
\fl \Phi=\left(
\begin{array}{cc}
\cos\eta \\
\sin\eta
\end{array} \right)f,\ppp \Phi_1=\left(
\begin{array}{cc}
\cos\eta \\
\sin\eta
\end{array} \right)f_1,
\end{eqnarray}
where $f$ and $f_1$ are arbitrary normalized functions from $L^2(\mathbb{R}^3)$.

Substituting these expressions into Eq. (\ref{12}), we reduce it to the form
\begin{eqnarray}\label{14}
\fl i\hbar \left\langle f_1\bigg|\frac{\partial f}{\partial t}\right\rangle= \langle f_1| H_{(+)} f\rangle = \langle
H_{(+)}f_1|f\rangle;
\end{eqnarray}
\begin{eqnarray}\label{15}
\fl H_{(+)}=\cos\eta\ooa H_{11}\cos\eta+\cos\eta\ooa H_{12}\sin\eta+\sin\eta\ooa H_{21}\cos\eta+\sin\eta\ooa
H_{22}\sin\eta.
\end{eqnarray}
Since $f_1$ and $f$ are arbitrary functions, it follows from the second equality in (\ref{14}) that the operator $H_{(+)}$
is the searched-for quantum-mechanical analogue of the operator $H$. From the first equality it follows that the function
$f$ must satisfy the equation
\begin{eqnarray} \label{16}
\fl i\hbar\frac{\partial f}{\partial t}=H_{(+)} f.
\end{eqnarray}
Taking into account Exps. (\ref{11}), as well as expressions $H_{11}=mc^2$, $H_{22}=-mc^2$ and $H_{12}=H_{21}=cp$, we
reduce (\ref{15}) to the form ($p$ commutes with $\eta$)
\begin{eqnarray} \label{17}
\fl H_{(+)}= c\sqrt{p^2+m^2c^2}.
\end{eqnarray}

Besides, if in Exp. (\ref{17}) we replace the operator $\eta$ with $\eta_{(-)}$ and take into account Exps. (\ref{11}), we
obtain a relativistic Hamiltonian bounded from above:
\begin{eqnarray*}
\fl H_{(-)}=-c\sqrt{p^2+m^2c^2}.
\end{eqnarray*}
This operator will not be considered further, as it lacks physical meaning.

Eq. (\ref{16}) is the desired quantum mechanical equation describing the single-component wave function $f$ of a spinless
boson. It is evident that its norm, as well as the average energy of the particle, are conserved:
\begin{eqnarray*}
\fl \langle f|f\rangle=\langle \Phi|\Phi\rangle=1,\ooo \langle f|H_{(+)}|f\rangle=const.
\end{eqnarray*}
In particular, in the $\mathbf{k}$-representation ($\mathbf{k}=\mathbf{p}/\hbar$) we have
\begin{eqnarray}\label{18}
\fl \langle f|f\rangle=\int_{\hat{\mathbb{R}}^3} |f(\mathbf{k},t)|^2 d^3\mathbf{k}=\int_{\hat{\mathbb{R}}^3}
\left(|\phi_1(\mathbf{k},t)|^2 + |\phi_2(\mathbf{k},t)|^2\right) d^3\mathbf{k}=1,\nonumber\\
\fl \langle f|H_{(+)}|f\rangle=c\int_{\hat{\mathbb{R}}^3} \sqrt{\hbar^2 k^2+m^2c^2}\ooa|f(\mathbf{k},t)|^2 d^3\mathbf{k},
\end{eqnarray}
where
\begin{eqnarray}\label{19}
\fl f(\mathbf{k},t)=\frac{1}{(2\pi)^{3/2}}\int_{\hat{\mathbb{R}}^3} f(\mathbf{r},t) e^{-i\mathbf{k}\mathbf{r}}
d^3\mathbf{k}.
\end{eqnarray}
Our next step is to establish, in the $\mathbf{k}$ representation, a connection between the wave function $f$ and the
solution $\Psi$ to the KGF equation.

\section{The relationship between the wave function and the solution to the Klein–Gordon–Fock equation} \label{link}

Note that, with taking into account Exps. (\ref{5}) for $\phi_1$ and $\phi_2$, Exps (\ref{1aa}) take the form
\begin{eqnarray}\label{32a}
\fl D_t\Psi=\epsilon e^{-i\eta}f,\ppp c(mc+ip)\Psi=\epsilon e^{i\eta}f.
\end{eqnarray}
Thus, taking into account expressions (\ref{9m})–(\ref{11}), from the first equation in Exps. (\ref{32a}) we obtain, in
the $k$-representation,
\begin{eqnarray}\label{33a}
\fl \Psi=\epsilon \frac{e^{-i\eta}}{D_t^{(+)}}f= \frac{\epsilon}{c^2} \frac{\sqrt{M}-i\sqrt{\mu}}{(M+\mu)^{3/2}}f.
\end{eqnarray}
In order to obtain the function $\Psi$ from the second equation in Exps. (\ref{32a}), we have firstly to note that
\begin{eqnarray*}
\fl (mc+ip)^{-1}=\frac{1}{c(m+2i\sqrt{M\mu})}= \frac{\left(\sqrt{M}-i\sqrt{\mu}\right)^2}{c(M+\mu)^2}.
\end{eqnarray*}
Thus,
\begin{eqnarray}\label{33b}
\fl \Psi=\epsilon \frac{e^{i\eta}}{c(mc+ip)}f= \frac{\epsilon}{c}
\frac{\sqrt{M}+i\sqrt{\mu}}{(M+\mu)^{1/2}}\cdot\frac{\left(\sqrt{M}-i\sqrt{\mu}\right)^2}{(M+\mu)^2} f= \frac{\epsilon}{c}
\frac{\sqrt{M}-i\sqrt{\mu}}{(M+\mu)^{3/2}}f.
\end{eqnarray}
As expected, Exps. (\ref{33a}) and (\ref{33b}) coincide, and we obtain an expression relating the functions $\Psi$ and
$f$:
\begin{eqnarray}\label{33d}
\fl \Psi=\frac{\epsilon}{c^2} g f\equiv  \frac{\epsilon}{c^2}\cdot \frac{\sqrt{M}-i\sqrt{\mu}}{(M+\mu)^{3/2}}\cdot f.
\end{eqnarray}
Thus, it remains to determine the constant $\epsilon$.

We proceed from the assumption that the quantum-mechanical expectation value of the boson energy $\langle
f|H_{(+)}|f\rangle$ coincides with the total energy of the scalar boson field $E_{tot}$ (see (\ref{305})). On the one
hand, taking into account Exps. (\ref{32a}) (as well as the fact that the quantity $E_{tot}$ also appears in the
Lagrangian (\ref{301})), from Exps. (\ref{304}) and (\ref{305}) we obtain that
\begin{eqnarray*}
\fl E_{tot}=\frac{1}{2E_{tot}} \int_{\hat{\mathbb{R}}^3} \left[ \left|D_t\Psi\right|^2+c^2\left|(mc+ip)\Psi\right|^2
\right]d^3\mathbf{k} =\frac{2\epsilon^2}{2E_{tot}} \int_{\hat{\mathbb{R}}^3}
\left|f(\mathbf{k},t)\right|^2d^3\mathbf{k}=\frac{\epsilon^2}{E_{tot}}.
\end{eqnarray*}
On the other hand, $E_{tot}=\langle f|H_{(+)}|f\rangle$. Thus, in the end, we arrive at the following:
\begin{eqnarray}\label{18m}
\fl \epsilon=E_{tot}=c\int_{\hat{\mathbb{R}}^3} \sqrt{\hbar^2 k^2+m^2c^2}\ooa\left|f(\mathbf{k},t)\right|^2 d^3\mathbf{k}.
\end{eqnarray}
Now that the normalization constant $\epsilon$ has been determined, we can proceed to define the momentum and orbital
angular momentum operators for a scalar boson (with zero electric charge and a mass $m$, which may be either non-zero or
zero).

\section{On the momentum and orbital angular momentum operators for a scalar boson} \label{operators}

In defining each of these two operators, we shall proceed from the premise that the quantum-mechanical expectation values
of the momentum and angular momentum operators for a scalar boson must coincide, respectively, with the total momentum and
total angular momentum of the scalar boson field.

\subsection{Momentum operator} \label{momentum}

Let $\hat{\mathbf{P}}_Q$ be the desired momentum operator, satisfying the equality
\begin{eqnarray}\label{70}
\fl \left\langle f|\hat{\mathbf{P}}_Q|f\right\rangle=\mathbf{P}_{tot}.
\end{eqnarray}
Taking into account Exps. (\ref{304}) and (\ref{305}), let's express $\mathbf{P}_{tot}$ via the wave function
$f(\mathbf{k},t)$:
\begin{eqnarray*}
\fl \mathbf{P}_{tot}=\frac{1}{2E_{tot}} \int_{\hat{\mathbb{R}}^3} \left[(D_t\Psi)(\mathbf{p}\Psi)^* +
(D_t\Psi)^*(\mathbf{p}\Psi)\right]d^3\mathbf{k} =\frac{2\epsilon^2}{2 E_{tot}c^2} \int_{\hat{\mathbb{R}}^3}
\frac{\mathbf{p}}{M+\mu}\left|f(\mathbf{k},t)\right|^2d^3\mathbf{k}\\ \fl =\frac{\epsilon}{c} \int_{\hat{\mathbb{R}}^3}
\frac{\hbar\mathbf{k}}{\sqrt{\hbar^2 k^2+m^2c^2}}\left|f(\mathbf{k},t)\right|^2d^3\mathbf{k}.
\end{eqnarray*}
Thus, it is obvious that equation (\ref{70}) is satisfied if
\begin{eqnarray}\label{71}
\fl \hat{\mathbf{P}}_Q=\frac{\epsilon}{c} \frac{\mathbf{p}}{\sqrt{p^2+m^2c^2}}.
\end{eqnarray}

\subsection{Orbital angular momentum operator} \label{orbit}

Let $\hat{\mathbf{M}}_Q$ be the desired orbital angular momentum operator for the scalar boson, satisfying the equality
\begin{eqnarray}\label{72}
\fl \left\langle f|\hat{\mathbf{M}}_Q|f\right\rangle=\mathbf{M}_{tot}\equiv\int_{\hat{\mathbb{R}}^3}
[\hat{\mathbf{r}}\times\mathbf{P}(\mathbf{k},t)] d^3\mathbf{k},
\end{eqnarray}
where $\hat{\mathbf{r}}=i\partial/\partial\mathbf{k}$. Let also $\hat{\mathbf{M}}$ be the usual angular momentum operator:
$\hat{\mathbf{M}}=[\hat{\mathbf{r}}\times\mathbf{p}]$. Now, taking into account expressions (\ref{304}) and (\ref{305}),
we can express $\mathbf{M}_{tot}$ in terms of the wave function $f(\mathbf{k},t)$:
\begin{eqnarray*}
\fl \mathbf{M}_{tot}=\frac{1}{2E_{tot}} \int_{\hat{\mathbb{R}}^3} \left[(D_t\Psi)(\hat{\mathbf{M}}\Psi)^* +
(D_t\Psi)^*(\hat{\mathbf{M}}\Psi)\right]d^3\mathbf{k}.
\end{eqnarray*}
Note that, according (\ref{33d}), $\hat{\mathbf{M}}\Psi=\epsilon(f\mathbf{M}g+g\mathbf{M}f)/c^2$. However, it easily to
check that $\mathbf{M}g=0$. Therefore,
\begin{eqnarray*}
\fl \mathbf{M}_{tot}=\frac{2\epsilon^2}{2 E_{tot}c^2} \int_{\hat{\mathbb{R}}^3}
\frac{f^*(\mathbf{k},t)\hat{\mathbf{M}}f(\mathbf{k},t)}{M+\mu}d^3\mathbf{k} =\frac{\epsilon}{c} \int_{\hat{\mathbb{R}}^3}
\frac{f^*(\mathbf{k},t)\hat{\mathbf{M}}f(\mathbf{k},t)}{\sqrt{p^2+m^2c^2}}d^3\mathbf{k}.
\end{eqnarray*}

Thus, it is obvious that equation (\ref{72}) is satisfied if
\begin{eqnarray}\label{71}
\fl \hat{\mathbf{M}}_Q=\frac{\epsilon}{c} \frac{\mathbf{M}}{\sqrt{p^2+m^2c^2}}.
\end{eqnarray}

\section{Solution of the quantum mechanical equation for a free relativistic KGF particle} \label{free}

In this section we present the general solution of Eq. (\ref{16}) for a free particle. That is we must find the normalized
general solution of the equation
\begin{eqnarray}\label{22}
\fl i\hbar \frac{\partial f(\mathbf{r},t)}{\partial t}= c\sqrt{\hat{p}^2+m^2c^2}f(\mathbf{r},t).
\end{eqnarray}
We will look for its solution in the form $f(\mathbf{r},t)=\frac{1}{(2\pi)^{3/2}}\int f(\mathbf{k},t)
e^{i\mathbf{k}\mathbf{r}} d^3 \mathbf{k}$ where the function $f(\mathbf{k},t)$ satisfies the equation
\begin{eqnarray}\label{23}
\fl i\hbar \frac{\partial f(\mathbf{k},t)}{\partial t}= c\sqrt{p^2+m^2c^2}f(\mathbf{k},t),
\end{eqnarray}
where $p$ is the multiplication operator $\hbar |\mathbf{k}|=\hbar k$.

Its normalized solution is
\begin{eqnarray*}
\fl f(\mathbf{k},t)= \mathcal{A}(\mathbf{k})e^{-iE(\mathbf{k}) t/\hbar}.
\end{eqnarray*}
where $E(\mathbf{k})=c\sqrt{\hbar^2 k^2+m^2c^2}$; $\mathcal{A}(\mathbf{k})$ is an arbitrary function belonging to the
Schwartz space and satisfying the normalization condition $\int_{\hat{\mathbf{R}}^3} |\mathcal{A}(\mathbf{k})|^2
d^3\mathbf{k}=1$. Thus,
\begin{eqnarray}\label{33dd}
\fl \Psi(\mathbf{k},t)= \frac{\epsilon}{c^2}\mathcal{A}(\mathbf{k})
\frac{\sqrt{M(k)}-i\sqrt{\mu(k)}}{\left[M(k)+\mu(k)\right]^{3/2}} \ooa e^{-iE(\mathbf{k}) t/\hbar}.
\end{eqnarray}
And, in the $\mathbf{r}$ representation, the general solution of the KGF equation (\ref{1}) is
\begin{eqnarray}\label{24a}
\fl \Psi(\mathbf{r},t)= \frac{1}{(2\pi)^{3/2}} \int_{\hat{\mathbf{R}}^3} \Psi(\mathbf{k},t) e^{i\mathbf{k}\mathbf{r}}
d^3\mathbf{k},
\end{eqnarray}
where $\epsilon$ is (see Exp. (\ref{18m})):
\begin{eqnarray*}
\fl \epsilon=c \int_{\hat{\mathbf{R}}^3} \sqrt{\hbar^2 k^2+m^2c^2}\ooa\left|\mathcal{A}(\mathbf{k})\right|^2
d^3\mathbf{k}.
\end{eqnarray*}
Note the (constant) norm of the function $\Psi$ does not equal to unit:
\begin{eqnarray*}
\fl \langle\Psi|\Psi\rangle=\frac{\epsilon^2}{c^2} \int_{\hat{\mathbf{R}}^3}
\frac{\left|\mathcal{A}(\mathbf{k})\right|^2d^3\mathbf{k}}{\hbar^2 k^2+m^2c^2} .
\end{eqnarray*}

\section{Conclusion}

The Klein-Gordon-Fock (KGF) theory is presented as a single-particle relativistic quantum theory. A Schrodinger-type
equation for a single-component wave function is derived. In addition to the KGF Hamiltonian, operators for momentum and
orbital angular momentum for a scalar boson are identified, which act within the space of single-component wave functions.
This theory is applicable to both massive and massless scalar bosons.





\end{document}